\documentclass[prl,twocolumn,showpacs,preprintnumbers,amsmath,amssymb]{revtex4-1}

\usepackage{graphicx}
\usepackage{dcolumn}
\usepackage{bm}

\begin{document}


\title{Skyrmion magnetic structure of an ordered FePt monolayer deposited on Pt(111)}

\author{S.~Polesya$^1$, S.~Mankovsky$^1$, S.~Bornemann$^1$, D.~K\"odderitzsch$^1$, J.~Min\'ar$^{1,2}$, 
        and  H.~Ebert$^1$}
\affiliation{1. Department Chemie/Phys. Chemie,
  Ludwig-Maximilians-Universit\"at M\"unchen, 81377 M\"unchen, Germany;\\
2. New Technologies - Research Center, University of West Bohemia,
Univerzitni 8, 306 14 Pilsen, Czech Republic }%

\date{\today}
             
\begin{abstract}
The effect of the Dzyaloshinsky-Moriya interaction on the magnetic
structure of an ordered FePt monolayer deposited on Pt (111) surface
has been investigated. In the ground state, the pronounced anisotropic
geometry of the FePt layer with alternating Fe and Pt chains gives rise
to a helimagnetic structure with a strong difference in the helicity
period  along the chains and perpendicular to  them. In the presence of 
an external magnetic field, the region of stable Skyrmion magnetic
structures in the $B-T$ phase diagram has been demonstrated
via Monte Carlo simulations using the parameters obtained within
first-principles electronic structure calculations. 
The present study demonstrates clearly that the ratio of the exchange
coupling parameters J/D for a deposited magnetic film - being of central
importance for the formation of Skyrmions - can be manipulated by 
growing an overlayer of 2-dimensional (2D) compounds with the atoms
carrying spontaneous magnetic 
moments separated by the atoms of non-magnetic elements.
\end{abstract}

\maketitle

\section{Introduction}

The novel topological magnetic structure called Skyrmion crystal (SkX),
observed recently in solids, attracts great interest due to various
promising physical properties, both from a academical and technological
point of view \cite{BR01,LKO+09,SRB+12,JMP+10}. This concerns, in
particular, the interaction
of Skyrmions (Sk) with an electric current leading to the
topological and anomalous Hall effect
\cite{NPB+09a,LKO+09,SRB+12,LKY+13} as well as giving potential access
to a new generation of spintronic devices based on current-driven Sk
manipulations \cite{JMP+10,LKY+13}. Predicted and investigated
theoretically by Bogdanov \emph{et al.}
\cite{BY89,BH94,RBP06,BLRB10,BHB+06,BHB+07,HBB08},   
 the SkX requires the presence of chiral interactions in the system, 
e.g., the spin-orbit coupling (SOC) induced
so-called Dzyaloshinsky-Moriya (DM) interaction intrinsic for 
systems with lack of inversion symmetry \cite{Dzy64}.   
This interaction can be responsible for a helimagnetic (HM) 
structure in the absence of an external magnetic field $\vec{B}_{ext}$, while the vortex-like Sk
magnetic structure can be stabilized by $\vec{B}_{ext}$ \cite{BY89,BH94,PLF+09,BLRB10,RLB10}.   
The presence of chiral interactions discerns
Skyrmions from so-called magnetic bubbles discussed in the literature
and appearing due to dipole-dipole interactions \cite{KBSR11}. 
The SkX state was observed experimentally for the first
time in the MnSi 
compound with B20 crystal structure \cite{MBJ+09,PLF+09}. Since then, the
SkX observation in bulk materials was reported also for others
systems, e.g., FeGe, Fe$_{1-x}$Co$_{x}$Si,
Mn$_{1-x}$Fe$_{x}$Si and Mn$_{1-x}$Co$_{x}$Si \cite{YOK+10a,YKO+11,BNF+10,MNA+10}. 
The conditions for Skyrmion stability in these systems were 
investigated theoretically by phenomenological considerations
based on the Landau-Ginzburg energy functional
\cite{PLF+09,BLRB10,RLB10,HZY+10}. 
To describe properly the itinerant-electron properties of magnetism in
these systems in the vicinity of the transition to the paramagnetic (PM)
state, Bogdanov and R\"o{\ss}ler account also for the energy related to
longitudinal spin fluctuations. 
Interesting is that in 3D materials the temperature window of SkX
stability in the phase diagram is rather narrow and located just below
the transition 
temperature to the PM state. In the case of 2D systems, the SkX
phase can exist in a larger temperature range approaching the
temperature $T = 0$~K \cite{YKO+11,HZY+10}.  

Note that SkX observation in 3D systems is rather difficult because of
the relatively weak DM interaction, while the formation of a 
non-collinear magnetic state requires the DM value strong 
enough to compete with the isotropic exchange interactions.
This requirement can be met in deposited monolayers as observed
experimentally in particular for Fe/Ir(111) 
overlayers \cite{BHB+06}, Mn/W(110) overlayers \cite{BHB+07} having
SOC-induced HM structure as well as Sk magnetic structure recently
observed in Fe(1ML)/Ir(111) \cite{HBM+11} and Pd/Fe(1ML)/Ir(111) \cite{RHM+13}. 

Despite the detailed theoretical
investigations based on a phenomenological approach,
a complete understanding of the conditions for the SkX appearance cannot 
be obtained without a detailed analysis of the microscopic origin of the
exchange interactions (both, isotropic and anisotropic) and
magneto-crystalline anisotropy (MCA), that needs a fully relativistic
\emph{ab-initio} description of the electronic structure 
of the systems under consideration.  Therefore theoretical schemes
based on relativistic density functional theory (DFT) \cite{MV79} are
applied to a large extend in computational simulations to support these
experimental efforts \cite{BHB+06, BHB+07,HBB08}.
In the present work we focus on the first-principles investigation of
the formation of the SkX state in a FePt ($2\times1$) monolayer deposited on
Pt(111), consisting of alternating Fe and Pt atomic chains with
             atoms occupying the Pt crystallographic positions.
 The geometry of the ($2\times1$) FePt/Pt(111) system is depicted 
in Fig.~\ref{fig:FePt_DM}~(a).
 
The magnetic properties of the ($2\times1$) FePt/Pt(111) clusters have
been recently studied both experimentally and theoretically
~\cite{HLK+09,MBM+09}, showing the presence of a pronounced
non-collinear magnetic structure in the system. 
This structure is governed by the strong DM interactions as well as
a specific atomic structure consisting of alternating Fe and Pt
chains leading to comparable values of isotropic and DM exchange
interactions between neighbouring Fe chains. Here we demonstrate that
the extension of the cluster to a monolayer leads to the formation of
the SkX state  in the presence of an external magnetic field.

\section{Computational details}

The $T$-$B$ phase diagram of FePt/Pt(111) has been
obtained performing Monte Carlo simulations using a
standard Metropolis algorithm \cite{Bin97} and based on the extended
Heisenberg model accounting for the SOC induced anisotropy of the
exchange interactions. The corresponding Hamiltonian is given by: 
\begin{eqnarray}
 H &=& 
-\sum_{i,j (i \neq j)} J_{ij}
 \hat{e}_i \cdot \hat{e}_j - \sum_{i,j (i \neq j)} \vec{D}_{ij}\cdot [\hat{e}_i\times
 \hat{e}_j] 
\nonumber \\
& & 
 +\sum_{i}  K_i (\hat{e}_i) - M \sum_{i} (\vec{B}_{ext} \cdot \hat{e}_i) \;.  
\label{Hspin_2}
\end{eqnarray}
with the isotropic exchange coupling parameters $J_{ij}$, the local
magnetic moment of atoms $\vec{M}_i = M\hat{e}_i$,
the DM vector $\vec{D}_{ij}$ and the anisotropy constants  $K_i(\hat{e}_i)$ accounting for the
on-site magnetocrystalline anisotropy (MCA) energy associated with each
individual moment oriented along $\hat{e}_i$. 

The main results presented here are obtained taking into account only the exchange
interactions between magnetic Fe atoms, neglecting the contribution from
the Pt atoms having a small induced magnetic moment which gives a small
correction to the critical temperature. 

A fully relativistic approach for the calculation of the exchange interaction
tensor $\underline{J}_{ij}$ based on the magnetic force theorem was
used \cite{USPW03}. This gives access to the isotropic and DM exchange
coupling parameters used in an extended Heisenberg model, Eq. (\ref{Hspin_2}).

The electronic structure calculations for a monolayer of FePt on
Pt(111), denoted FePt/Pt(111), have been performed
within the local density approximation for DFT \cite{VWN80}, using the
spin-polarized relativistic Korringa-Kohn-Rostoker multiple scattering
formalism~\cite{Ebe00}.
 In this scheme, the Dirac Green function was
calculated self-consistently for FePt/Pt(111) assuming pseudomorphic
deposition on a Pt slab consisting of 37 atomic layers and having the
experimental lattice constant of bulk Pt ($a=3.924$~\AA{}). 
All calculations have been 
performed within the atomic sphere approximation (ASA) to the potentials
and lattice relaxations near the surface have been neglected (see
Ref.~\cite{BSM+12} for more details).

\section{DM interactions}

\begin{figure}[b]
\includegraphics[width=0.3\textwidth,angle=0,clip]{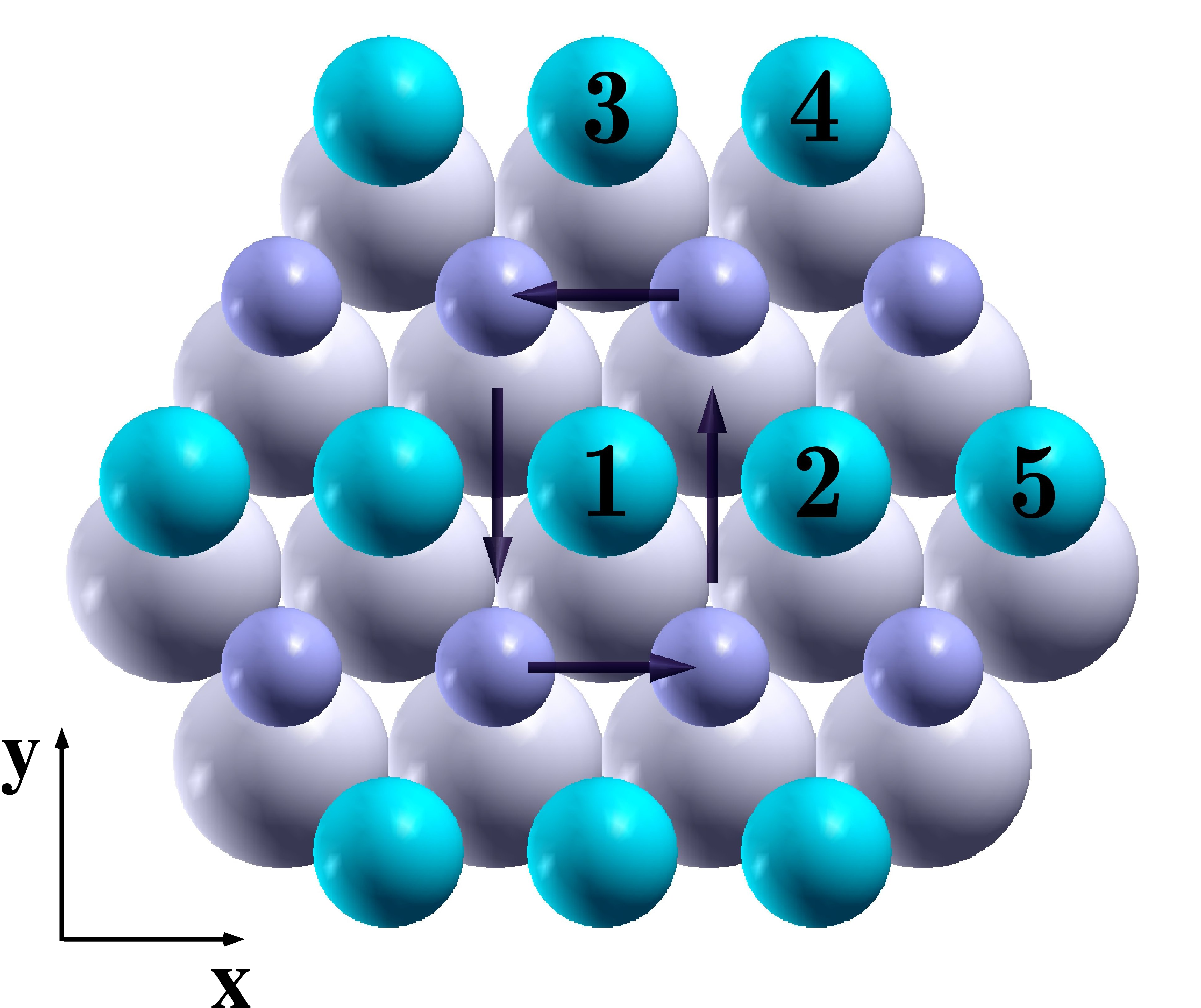}\;(a)
\includegraphics[width=0.4\textwidth,angle=0,clip]{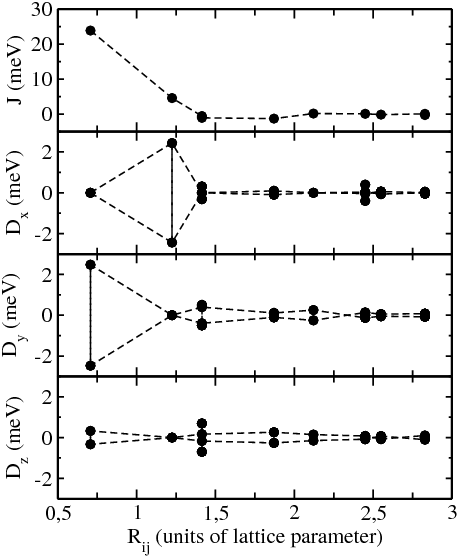}\;(b)
\caption{\label{fig:FePt_DM} (a) Geometry of the system and directions of
  in-plane components of DM exchange interactions (shown by the arrows) between Fe1 and Fe2
  atoms, Fe1 and Fe3 atoms, etc. (b)  Calculated exchange coupling
  parameters: isotropic, $J_{ij}$, for Fe-Fe (circles) and Fe-Pt
  (diamonds) (top panel), and $D^x_{ij}$, 
  $D^y_{ij}$ and $D^z_{ij}$ components of DM interactions between Fe
  atoms for FePt/Pt(111).  }  
\end{figure}

The structure of the ($2\times1$) FePt/Pt(111) system is depicted 
in Fig.~\ref{fig:FePt_DM}~(a) consisting of a monolayer of alternating
Fe and Pt atomic chains deposited on a Pt(111) surface.
The calculated magnetic moments of the Fe and Pt atoms within the
  FePt monolayer are $~3.14 \mu_B$, and $~0.22 \mu_B$, respectively. The
  induced magnetic moment in the Pt substrate decreases quickly away
  from the surface having $~0.18 \mu_B$ and $~0.03 \mu_B$ in the
  interface and next to the interface layers, respectively. 
Figure \ref{fig:FePt_DM}~(b) represents the isotropic exchange coupling
parameters $J_{ij}$ and $D^x_{ij}, D^y_{ij}$ and $D^z_{ij}$ components
of the DM interaction vector.       
These parameters for the FePt film on Pt(111) are very close to 
	those obtained for the FePt/Pt(111) cluster \cite{MBM+09} (Note
	however the different form of the Heisenberg Hamiltonian used in
	\cite {MBM+09}, that leads to the exchange parameters being two
	times bigger per definition than those used in the present work).
Strong ${\vec D}_{ij}$ interactions between the first neighboring Fe
atoms indicate the favorite conditions for the appearance of a HM
structure in the system. The properties of ${\vec D}_{ij}$  along
different directions ${\vec R}_{ij}$ (Fig. \ref{fig:FePt_DM}~(b) and
Table \ref{one} are clearly governed by 
the system symmetry \cite{Mor60a, CL98}. The symmetry plane between
the in-chain Fe atoms 1 and 2 (see Fig.\ \ref{fig:FePt_DM}~(a)) forces the
$\vec{D}_{12}$ component along the chain to be equal to 0. The symmetry plane
crossing the Fe atoms at positions 1 and 3 allows a non-zero component of
$\vec{D}_{13}$ interaction along the direction parallel to Fe and Pt chains.
Summarizing the results for the DM interaction for other distances one can conclude
that they have pronounced in-plane components (see Fig.\
\ref{fig:FePt_DM}) responsible for a rotation of the magnetic moments within
a corresponding plane orthogonal to the film. As it follows from the model
consideration by Fert and Levy \cite{FL80} giving the expression for DM
interactions in the form ${\vec D}_{ij} \sim \frac{({\vec R}_{in} \cdot {\vec
    R}_{nj}) [{\vec R}_{in} \times {\vec R}_{nj}]}{R_{ij}R_{in}R_{nj}}$
($j,i$ correspond to the sites of magnetic atoms while 
$n$ corresponds to the sites of atoms mediating exchange interaction),
the DM interactions are mediated by the Pt atoms and
their big magnitude is essentially determined by strong SOC of the Pt
atoms. 
This allows also to draw conclusions about the small contribution
         to the $D^z$ component of the DM vector at short distances
         $R_{ij}$, coming (i) from the substrate Pt atoms
         as well as (ii) from the Pt atoms within the FePt monolayer
         (this contribution is non-zero only in the presence of the
         Pt substrate responsible for the breaking of the inversion
         symmetry).
On the other hand, the large magnitude of the $D^x$ and $D^y$
components is fully determined by the substrate Pt atoms, giving
evidence to the crucial role of their strong SOC values for the 
in-plane components of the $\vec{D}_{ij}$ interactions between the Fe atoms.  
At large distances $R_{ij}$ all three components have the same
order of magnitude, due to contributions of many Pt atoms involved into
the mediation of Fe-Fe DM exchange interactions. 
Thus, one can clearly see a pronounced SOC induced effect in the
present system leading to DM couplings strong enough to
compete with the isotropic exchange coupling and to create a pronounced
non-collinear magnetic structure.

\begin{table}\caption{\label{one} Components of the DM
    vector $\vec{D}_{ij}$ and the isotropic exchange constant $J_{ij}$
    (in meV) for the FePt/Pt(111).} 
\begin{center}
\begin{tabular}{l||c|c|c|c|c|c}
\hline
$i$-$j$ & \hspace{0.1mm} $R_{ij}$\hspace{0.1mm} &   \hspace{0.1mm}
$D^x_{ij}$\hspace{0.1mm} &\hspace{0.1mm} $D^y_{ij}$\hspace{0.1mm}
&\hspace{0.1mm} $D^z_{ij}$\hspace{0.1mm} &  $J_{ij}$ & \\ 
\hline
 $1$-$2$ & 0.707 &0.00 & 2.44 & 0.39 & 23.90  &    \\ 
\hline
 $1$-$3$ & 1.225 &-2.47 & 0.00 & 0.00 & 4.59 &    \\ 
\hline
 $1$-$4$ & 1.414 & -0.31 & 0.50 & -0.69 & -0.56 &    \\ 
\hline
 $1$-$5$ & 1.414 & 0.00  & 0.39 & -0.17 & -1.07 &    \\ 
\hline
\end{tabular}
\end{center}
\end{table}

\section{Helimagnetic structure}

\begin{figure}[b]
\includegraphics[width=0.5\textwidth,angle=0,clip]{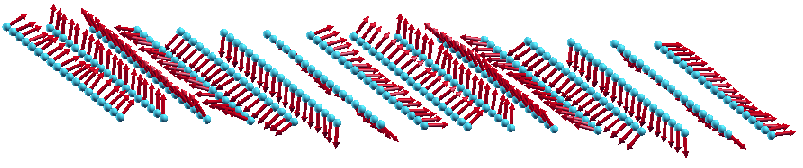}\;
\caption{\label{fig:FePt_HM} Helimagnetic structure in FePt/Pt(111). The arrows
  represent the magnetic moments of Fe atoms.}  
\end{figure}

Magnetic torque calculations show a rather strong in-plane MCA of $1.1$
meV per Fe atom with the magnetic easy axis being  perpendicular to the
Fe and  Pt chains. The MCA energy was taken into account in all present MC 
simulations based on Heisenberg Hamiltonian Eq.~(\ref{Hspin_2}). 
First, the calculations have been performed in the absence of an
external magnetic field. 
The resulting non-collinear magnetic structure obtained at 
$T = 1$~K is presented in Fig.\ \ref{fig:FePt_HM}. The period of the helicity in
the continual model represented by Landau-Ginzburg energy functional
is determined by the ratio of exchange stiffness and Dzyaloshinsky-Moriya
constants, $A/D$ \cite{BH94}. In general, these values are represented by
tensors of first rank, $D^{\alpha} = |\sum_{j} \vec
D_{0j}R^{\alpha}_{0j}|$, and  second rank, $A^{\alpha\beta} = \sum_{j}
J_{0j}R^{\alpha}_{0j}R^{\beta}_{0j}$ \cite{LKG84}, characterizing a spatial   
anisotropy of the system, that is important for the present case having
$C_{s}$ symmetry. In particular, these values give the 
anisotropy of the energy of spin-wave excitations with wave vector $\vec q =
q\hat{n}_\alpha$ along different directions, $\hat{n}_\alpha = \hat{x},
\hat{y}$: $\omega(q^{\alpha}) = {\tilde A}^{\alpha\alpha}(q_{\alpha})^2 +
{\tilde D}^{\alpha}q_{\alpha} + \omega_0$, with spin-wave coefficients
${\tilde A}^{\alpha\alpha} = \frac{2\mu_B}{M} A^{\alpha\alpha} $ and 
${\tilde D}^{\alpha} = \frac{2\mu_B}{M} D^{\alpha} $, and $\omega_0$ 
the spin-wave gap created by magnetocrystalline anisotropy.
%
%
The strong anisotropy of the exchange stiffness tensor
is well recognized considering the first-neighbor isotropic exchange
coupling parameters (see Table~\ref{one}) within and between the Fe chains.
This leads to a complex HM structure with
a different period along different directions, i.e. the period along the
chains is essentially longer than perpendicular to the Fe chains, as
can be clearly seen in Fig.\ \ref{fig:FePt_HM}. 
Raising the temperature from $T = 0$~K, the magnetization of the system
exhibits two phase transition: from HM to FM state at $T_H = 40$~K, and
from FM to the paramagnetic (PM) at $T_C = 90$~K.

\begin{figure}[b]
\includegraphics[width=0.2\textwidth,angle=0,clip]{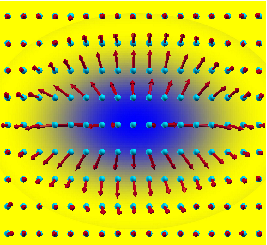}\;(a)
\includegraphics[width=0.2\textwidth,angle=0,clip]{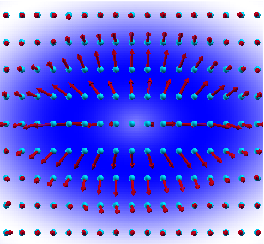}\;(b)
\includegraphics[width=0.2\textwidth,angle=0,clip]{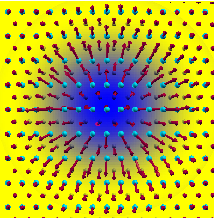}\;(c)
\includegraphics[width=0.2\textwidth,angle=0,clip]{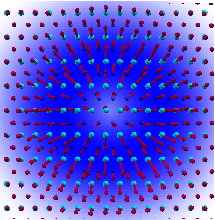}\;(d)
\caption{\label{fig:Skyrm} Magnetic moment distribution within the
  Skyrmion. Yellow  and blue colors in (a) represent schematically the
  region giving 
  gain and loss of Zeeman energy in the presence of a magnetic field; blue
color in (b) shows the region giving loss of the exchange energy
contributed by in-plane components of magnetic moments. (c) and (d):
Structure of single Sk obtained with the contributions of the Fe-Pt exchange
interactions taken into account (the same color-code as in (a) and
(b)). Long arrows show the spontaneous magnetic moments on Fe atoms,
short arrows indicate the induced magnetic moments on Pt atoms.}   
\end{figure}

\section{Skyrmion magnetic structure}

The presence of an external magnetic field $\vec{B}_{ext}$ perpendicular
to the surface plane, exceeding a certain critical value, gives rise to
the formation of Skyrmions in the system under consideration. 
The distribution of the magnetic moments within the single Skyrmion obtained
using the MC simulation is shown in Fig.\ \ref{fig:Skyrm}.
The topological quantum number (or winding number) calculated for such a
 magnetic texture is equal to -1 (see Appendix), being a property of
 single Skyrmion (see, e.g.,\cite{MBJ+09,OOS+12})
One
can clearly see the rather small Skyrmion size (compared with the period
of the HM structure) due to a small $A/D$ ratio. Its different size in two
orthogonal directions within the plane is determined by the spatial
anisotropy of the exchange coupling parameters, $J_{ij}$ and ${\vec
  D}_{ij}$. The in-plane tangential components of the magnetic moments
governed by $D^z_{ij}$ component of DM interactions are close to $0$, 
in line with the discussions by R\"o{\ss}ler et al. \cite{RLB10} on the
magnetic moment distribution within the Skyrmions in FM system with
$C_{nv}$ symmetry.  

Analyzing the requirements for the Skyrmion formation for FePt/Pt(111),
one can use rather simple qualitative arguments. In this consideration
the effect of a demagnetizing field can be neglected being small in the
case of a magnetic monoatomic overlayer.  
A competition of isotropic exchange and DM
interactions determine the period of the HM structure without magnetic
field and anisotropy by minimizing the energy of the system. 
In the presence of an out-of-plane magnetic field the minimization of the Zeeman
energy is realized by minimizing the 2D area with magnetic moments 
opposite to the direction of the magnetic field (the Zeeman energy gain is
shown by yellow color, while the energy loss is shown by blue in 
Fig.\ \ref{fig:Skyrm}~(a)). However, this costs the 
energy originating from the inter-atomic isotropic exchange, due to
a non-collinear orientation of the in-plane components of the magnetic
moments within the Skyrmion (see Fig.\ \ref{fig:Skyrm}~(b), where the
exchange energy loss is shown by blue color), assuming that the Skyrmion
size is close to the period of the HM structure. The competition of
these two factors determine the 
conditions for the formation of Skyrmions at some critical value of the
external magnetic field $\vec{B}_{ext}$. Despite the strong
simplification neglecting the 
role of other contributions (due to exchange interactions and MCA) to
the SkX energetics, this qualitative consideration visualizes the role
of the magnetic field for the SkX formation as a ground state ($T = 0$~K),  
relevant in the case 2D system. 

As was mentioned above, the 2D anisotropy of the system leads to a
corresponding shape of the Skyrmions obtained via calculations accounting for
Fe-Fe exchange interactions only, 
 as soon as only Fe atoms carry spontaneous magnetic moments weakly
  dependent on magnetic configuration. Since isotropic exchange
  interactions between Fe atoms within the 
  Fe chains are much stronger than the inter-chain Fe-Fe interactions,
  their competition with DM interactions results in the weaker
  modulation of magnetic structure along the chains when compared to the
  direction perpendicular to the chains. However, in systems containing 
  atoms with induced magnetic moments, i.e., Pt in the present case,
  the Fe-Pt interactions, $J^{Fe-Pt}_{ij}$, can lead to an additional
  contribution to the 
  Fe-Fe exchange energy if $J^{Fe-Pt}_{ij}$ (shown in
  Fig. \ref{fig:FePt_DM}) value is not negligible (see 
  discussions, e.g. \cite{Mry05,LME+06,PMS+10}). However, to account for
  this contribution in MC calculations is not straight-forward and is
  rather time consuming.  
 Therefore, we have performed here the calculations accounting for
 Fe-Pt interactions only to investigate their effect on the shape of the
 Skyrmion, following the scheme described in Ref. \cite{PMS+10}.
Based on this, we conclude that the Pt contribution could
 modify the phase diagram, but these changes should not be crucial. 
The result, obtained at $T = 1$~K, is presented in
Fig.\ \ref{fig:Skyrm}~(c). One can see that the Skyrmion shape in this case
is more symmetric. This occurs due to an additional Fe-Pt exchange
contribution competing with DM interactions and resulting in the
increase of the effective J/D (see the above discussion) ratio along
the direction perpendicular to the Fe chains, and leading that way to the
increase of the Skyrmion size along the
direction perpendicular to the Fe chains.

\section{Phase diagram}

\begin{figure}[b]
\includegraphics[width=0.45\textwidth,angle=0,clip]{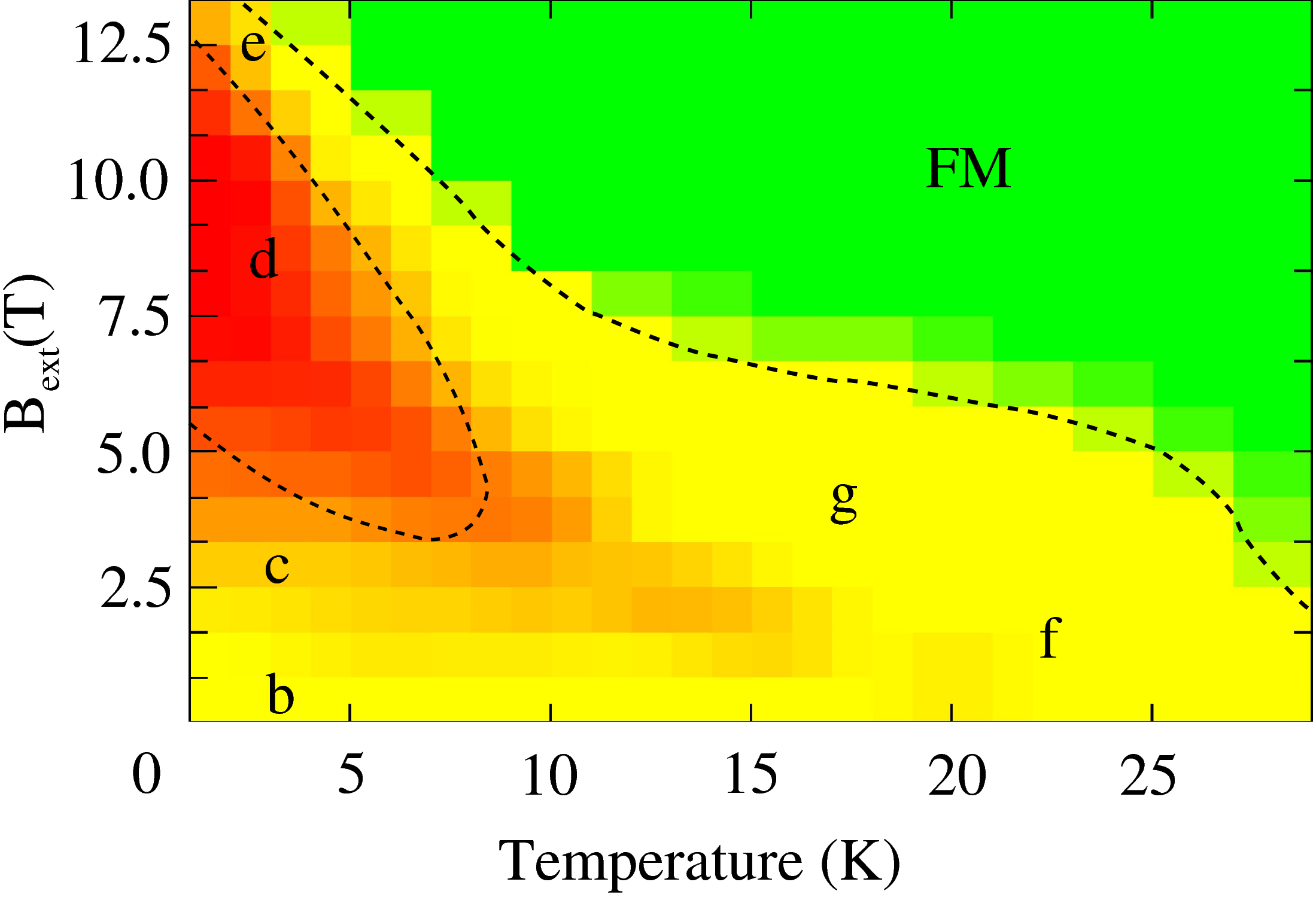}\;(a)
\includegraphics[width=0.128\textwidth,angle=0,clip]{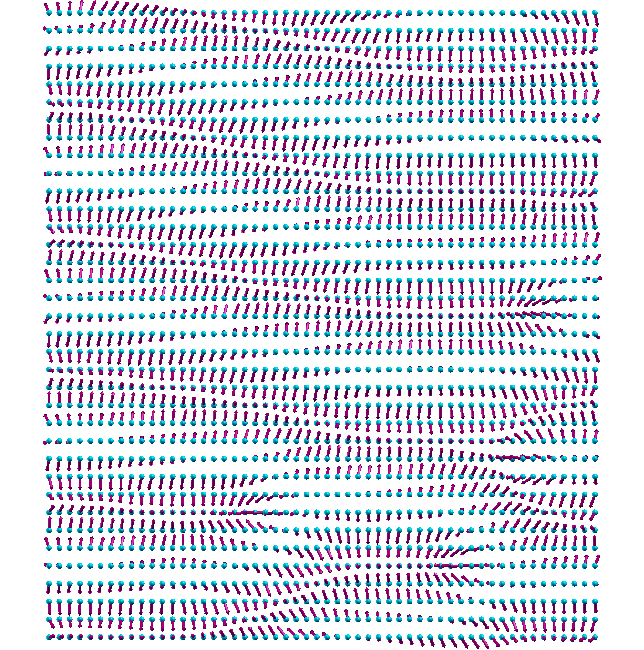}\;(b)
\includegraphics[width=0.128\textwidth,angle=0,clip]{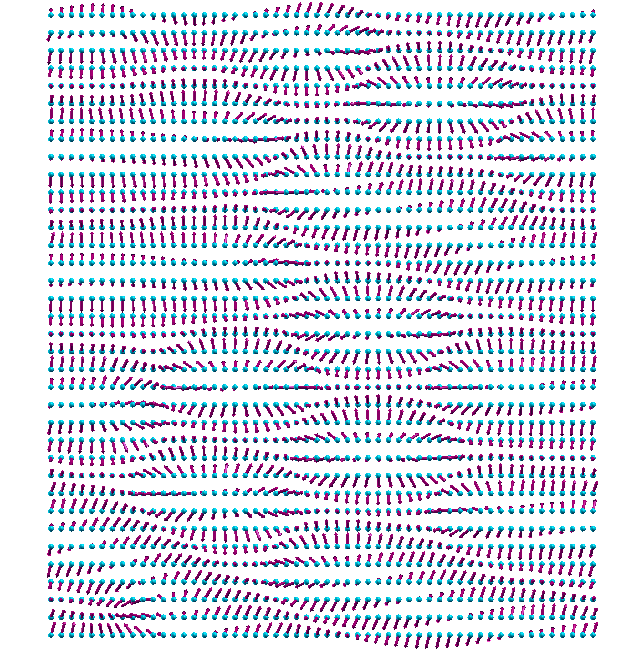}\;(c)
\includegraphics[width=0.128\textwidth,angle=0,clip]{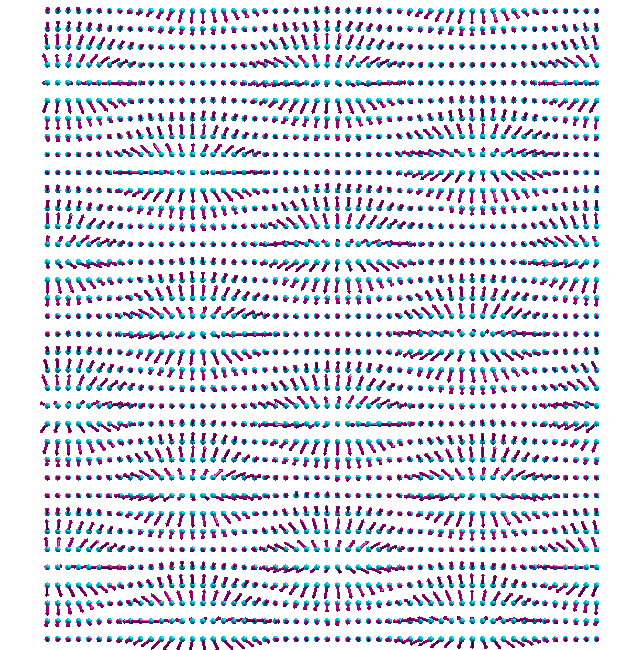}\;(d)
\\
\includegraphics[width=0.13\textwidth,angle=0,clip]{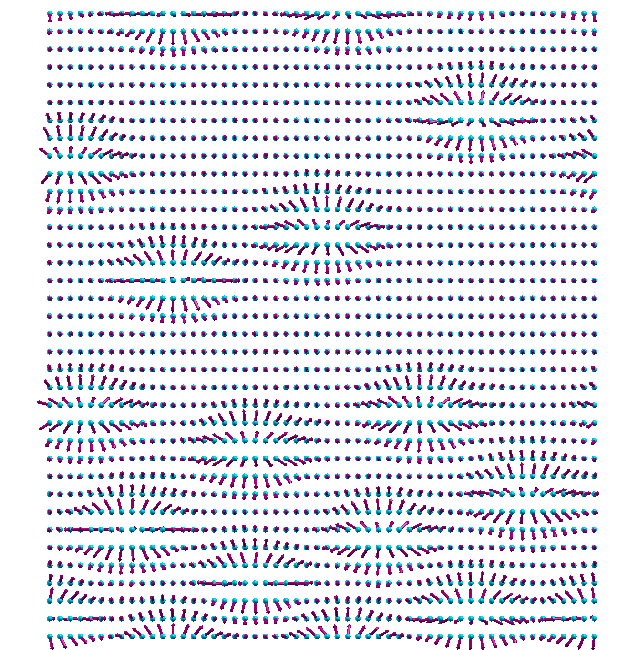}\;(e)
\includegraphics[width=0.13\textwidth,angle=0,clip]{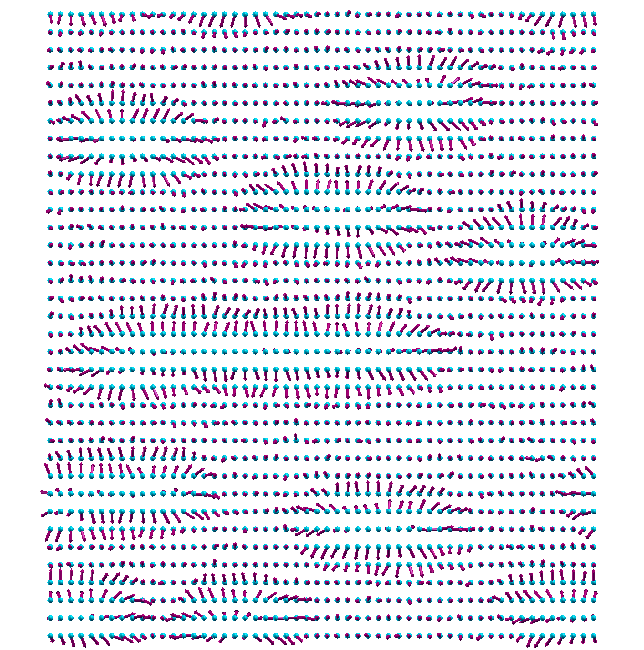}\;(f)
\includegraphics[width=0.13\textwidth,angle=0,clip]{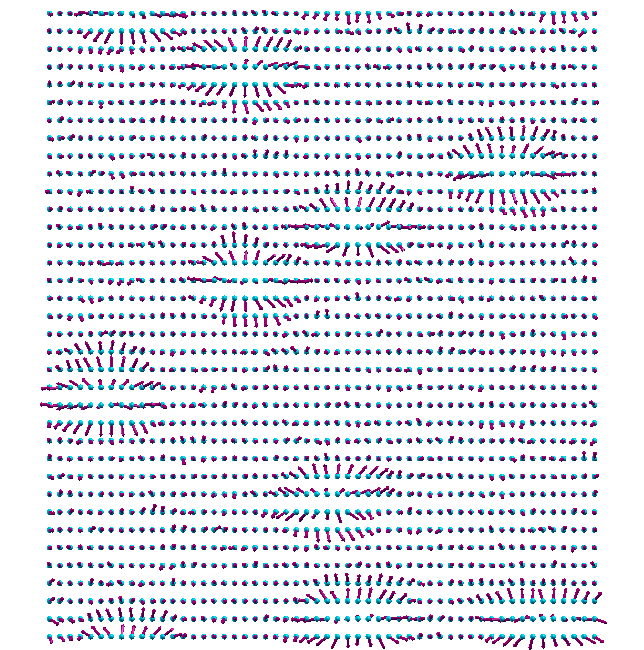}\;(g)
\caption{\label{fig:phase_diagr} 
(a) Low-temperature part of $B$-$T$ phase diagram calculated for
FePt/Pt(111); (b)-(g) representative magnetic structures of the phase
diagram regions indicated in (a) obtained at $B = 0.0$~T, $T = 3.0$~K (b),
$B = 2.5$~T, $T = 3.0$~K (c), $B = 7.5$~T, $T = 3.0$~K (d), $B =
12.5$~T, $T = 3.0$~K (e), $B = 1.0$~T, $T = 22.0$~K (f), $B = 3.5$~T, $T
= 17.0$~K (g) }  
\end{figure}

The magnetic structure of FePt/Pt(111) and its behavior at
different temperature and external  
magnetic field, perpendicular to the surface, has also been investigated
via Monte Carlo simulations, using the standard Metropolis
  algorithm. For this, only the interactions within the Fe
subsystem with well defined local magnetic moments have been taken into
account. The MC unit cell containing 2500 Fe atoms was extended using
periodic boundary conditions. The in-plane magnetic anisotropy was taken
into account with MCA direction perpendicular to Fe chains and $E_{MCA}
= 1.1$~meV (obtained in present calculations). The important role of the
magnetic anisotropy for the stabilization of SkX state in some systems was
discussed in the literature \cite{BLRB10,RLB11}. Therefore, a set of
additional MC calculations has been performed for the out-of-plane MCA
direction (with the MCA energy equal to 1.1 meV) as well as 
for the MCA energy set to 0. All these results exhibit rather
small differences indicating a weak effect of the MCA and therefore
the main responsibility of the exchange interactions (isotropic and DM)
for the SkX stabilization in the system under consideration. Therefore,
we present below only the results obtained for in-plane MCA found in the
present DFT calculations.

\begin{figure}[b]
\includegraphics[width=0.4\textwidth,angle=0,clip]{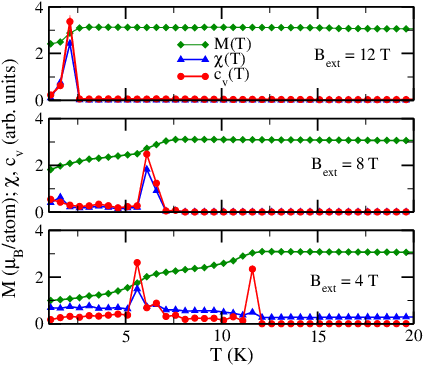}\;
\caption{\label{fig:capa} 
Temperature dependent magnetization $M(T)$ (diamond), susceptibility $\chi(T)$ (triangles) and 
heat capacity $c_v(T)$ functions represented for three different
magnetic fields: $B = 12$~T (top panel), $B = 8$~T (middle panel) and
$B = 4$~T (bottom panel). }  
\end{figure}

More results of the MC simulation can be seen in Fig.\
\ref{fig:phase_diagr} representing the low-temperature part of the
$T$-$B$ phase diagram. 
The critical temperatures within the MC simulations have been obtained
from an analysis of the heat capacity $c_v(T,B)$ and susceptibility
$\chi(T,B)$. Figure \ref{fig:capa} shows the
temperature dependence of these parameters for three different values
of the magnetic field:  $B = 12$~T (top panel), $B = 8$~T (middle panel) and
$B = 4$~T (bottom panel). In the first two cases the $c_v(T)$ curves
have only one maximum corresponding to the boundary of the SkX region,
while the two maxima of $c_v(T)$  in the case of $B = 4$~T are associated with
the low- and high-temperature boundaries of the SkX region shown in  Fig.\
\ref{fig:phase_diagr} in red. To improve the
statistical error a series of independent calculations have been
performed here for each point at the 
phase diagram: (i) by varying the temperature $T$ at a fixed magnetic
field $B$ value and (ii) by varying the magnetic field, keeping the
temperature constant. Nevertheless, one has to admit that perfect phase
boundaries have not been obtained. As soon as the density of Skyrmions
reproduces the SkX phase boundaries with the same accuracy as
$c_v(T,B)$ and $\chi(T,B)$, the phase diagram is plotted representing
the density of Skyrmions at each $(B,T)$ point. 
  Without magnetic field the system exhibits a rather pronounced
  SOC-induced HM structure (Fig.\ \ref{fig:phase_diagr}~(b)) below the
  critical temperature $T_H = 40$~K. The FM structure is stabilized in
  the temperature region from $T_H = 40$~K to $T_C = 90$~K, while the
  following temperature increase drives the system into the PM state. The
  critical temperatures have been evaluated from the temperature
  dependent behaviour of the magnetic susceptibility, using the average
  magnetization in the system as an order parameter. 
An external magnetic field, $\vec{B}_{ext}$,
leads at low temperature to the formation of Skyrmions coexisting with
the HM structure (Fig.\ \ref{fig:phase_diagr}~(c)).  
Raising of the magnetic field at low temperature leads to the
formation of a Skyrmionic lattice (Fig.\ \ref{fig:phase_diagr}~(d)). 
  The density of Skyrmions is shown in the phase diagram by red and
  yellow, representing the SkX area with highest Skyrmion density by
  red, while the wide region with low Skyrmion density is colored in
  yellow. Note that the boundary between the low Skyrmion (yellow) and
  FM (green) phases was plotted using only the data on the temperature
  dependence of Skyrmion density which is very low in this region. The
  dashed lines are drawn as a guide for the eyes to visualize the
  approximate boundaries separating different phases. 
At high magnetic fields and low temperature the concentration of 
Skyrmions is reduced resulting in the mixed FM + Sk state (Fig.\
\ref{fig:phase_diagr}~(e)) until the transition to the FM state.
At these values of the magnetic field the transition to the FM + HM + Sk
state does not occur.
On the other hand, at high temperature and low magnetic field, as it was
discussed above, a mixed FM + HM + Sk state is observed (Fig.\
\ref{fig:phase_diagr}~(f)), while an increasing of the magnetic field leads to a
mixed FM + Sk state (Fig.\ \ref{fig:phase_diagr}~(g)). 
At all values of the magnetic field the temperature increase leads to a
transition from the non-collinear magnetic state to the FM state.  

Thus, the behavior of the magnetic structure of FePt/Pt(111)
in the presence of a finite external field shows striking differences when
compared with the properties of 3D systems, e.g. FeGe, MnSi, MnGe, etc.
This is, first of all, due to the well defined Fe local magnetic 
moment indicating a small effect of 
longitudinal fluctuations on the formation of Skyrmionic magnetic
properties. A Skyrmion lattice state in the phase diagram appears at low
temperature, in contrast to  3D systems, where the Skyrmion lattice formation
occurs close to the temperature of the transition to the disordered magnetic
state. Because the Curie temperature in the system is rather high,
the temperature increase leads first to the transition from the Skyrmionic
to the FM state and then from the FM to the PM state, in contrast to the SkX-PM
transition observed in the 3D bulk compounds FeGe, MnSi and MnGe. 
Finally, we would like to stress that strong in-plane components of the 
$\vec{D}_{ij}$ interactions, governed by the substrate Pt atoms in FePt/Pt(111), lead
to a small Sk size (see above), that makes this system
attractive for technological applications. Note however,
that too strong DM interactions can lead to a stabilization of the
SkX as a ground state \cite{HBM+11,LLZ11}, even without external magnetic field.
One can expect different behavior of this state when compared to those
discussed above, although detailed investigations of its phase diagram
so far have not been done.

\section{Acknowledgements}

Financial support by the DFG via SFB 689 (Spinph\"anomene in reduzierten
Dimensionen) is thankfully acknowledged.  

\section{Appendix}

To calculate the winding number for an isolated Skyrmion obtained within MC
calculations, its magnetic configuration $\hat{n}(\vec{R}_i) =
\vec{m}(\vec{R}_i)/|\vec{m}(\vec{R}_i)| $ can be
represented by a continuous model in cylindrical coordinates.
In this case, the direction of the magnetic moment is equal to
$\hat{n}(\vec{r}) =
(|\vec{n}_{||}|\mbox{cos}\phi,|\vec{n}_{||}|\mbox{sin}\phi),|\vec{n}_{z}|) 
= (\mbox{sin}\theta\mbox{cos}\phi ,\mbox{sin}\theta\mbox{sin}\phi
,\mbox{cos}\theta) $ for each point 
$\vec{r} = (\rho \mbox{cos}\phi,\rho \mbox{sin}\phi)$ in 
the $xy$ (film) plane. Note that the in-plane component of the magnetic
moment is always aligned along the radius $\hat{\rho}$: $\vec{n}_{||} =
(|\vec{n}_{||}(\rho)|\mbox{cos}\phi,|\vec{n}_{||}(\rho)|\mbox{sin}\phi) || (\rho \mbox{cos}\phi,\rho
\mbox{sin}\phi) $, and
$\theta (\rho)$ is an angle with respect to the $z$ axis, depending only on
the distance from the center of the Skyrmion.
 In this case the winding number can be calculated by
doing the necessary integration using polar coordinates:

\begin{eqnarray}
 W &=& \frac{1}{4\pi} \int dx \int dy
 \hat{n} \cdot  \left[\frac{\partial\hat{n}}{\partial{x}} \times \frac{\partial\hat{n}}{\partial{y}}\right] 
 \; \nonumber \\
 &=& \frac{1}{4\pi} \int_{0}^{R_s} \rho d\rho \int_{0}^{2\pi} d\phi
 \left[\frac{\partial\hat{n}}{\partial{x}} \times
   \frac{\partial\hat{n}}{\partial{y}}\right] \;.
\label{App_1}
\end{eqnarray}

Taking into account the expressions for the derivatives in polar
coordinates (see, e.g., \cite{VMK88})
%
%
and assuming that $\theta(r=0) = \pi$ and $\theta(r=R_s) = 0$ (where
$R_s$ is the Skyrmion radius) one obtains the winding number according
to the shape of the Skyrmion:    

\begin{eqnarray}
 W &=& \frac{1}{4\pi} \int_{0}^{R_s} \rho d\rho
 \left[\frac{1}{\rho}\frac{\partial\theta}{\partial{\rho}}
   \mbox{sin}\theta \right]
 \int_{0}^{2\pi} d\phi = -1
 \;.
\label{App_3}
\end{eqnarray}


\end{document}